\documentclass[pre,twocolumn,floatfix,showpacs]{revtex4}
\usepackage{graphicx,amsmath,amssymb,mathptm}
\newcommand{\reg}{\mathcal{R}}
\begin{document}
\title{Broad edge of chaos in strongly heterogeneous Boolean networks} 
\author{Deok-Sun Lee}
\affiliation{Center for Complex Network Research and Department of Physics, 
  Northeastern University, Boston, MA 02115, USA}
\author{Heiko Rieger}
\affiliation{Theoretische Physik, Universit\"{a}t des Saarlandes, 
  66041 Saarbr\"{u}cken, Germany}
\date{\today}
\begin{abstract}
The dynamic stability of the Boolean networks representing a model for 
the gene transcriptional regulation (Kauffman model) is studied by 
calculating analytically and numerically the Hamming distance between 
two evolving configurations. This turns out to behave in a universal way 
close to the phase boundary only for in-degree distributions with a finite 
second moment. In-degree distributions of the form $P_d(k)\sim k^{-\gamma}$ 
with $2<\gamma<3$, thus having a diverging second moment, lead to a slower 
increase of the Hamming distance when moving towards the unstable phase 
and to a broadening of the phase boundary for finite $N$ with decreasing 
$\gamma$. We conclude that the heterogeneous regulatory network connectivity 
facilitates the balancing between robustness and evolvability in living 
organisms.
\end{abstract}
\maketitle

\section{Introduction}
\label{sec:intro}

Complete genome sequencing and the analysis of the binding of transcriptional 
regulators to specific promoter sequences have uncovered the global organization 
of the gene transcriptional regulatory  network in well-studied organisms such 
like {\it Escherichia coli}~\cite{thieffry98} and yeast 
{\it Saccharomyces cerevisiae}~\cite{tilee02}. The gene network describes 
a directed relationship - regulation -
between different genes, and its architecture is characterized by broad 
connectivity distributions ~\cite{thieffry98,tilee02,dobrin04,guelzim02}, 
over-representation of selected motifs ~\cite{shenorr02}, and so on. These 
features are rarely found in random networks, probably being the consequence 
of evolutionary selection. Therefore illuminating the functional characteristics 
associated with those discovered structural features can help trace back their origins. 
In this work, we show heterogeneous connectivity can facilitate 
the balancing between dynamical stability and instability. Both robustness and 
evolvability are essential for living organisms, which achieve their specific 
phenotype by their gene expression program~\cite{babu04}. Thus the 
transcriptional regulatory network should be organized in a way that supports 
the coexistence of these apparently contradictory properties and from this 
perspective, it has been proposed that the gene network should be at the 
boundary between stable and unstable phases, called the edge of chaos
~\cite{kauffman}. The question then arises: What are the characteristics of 
the network architecture that can support the requirement to be located at 
the edge of chaos? A simple model incorporating recently available information 
turns out to be useful to answer this question.

The Kauffman model~\cite{kauffman} was used in the past to study the gene network
dynamics which is far from completely known because of its complexity. In this 
model, each node has a Boolean variable, $1$ or $0$, the discretized expression 
level, evolving regulated by other $K$ nodes according to the quenched rules 
that are randomly distributed with a parameter $p$. In spite of these 
simplifications involved, the model revealed detailed relations between the 
dynamical stability against perturbations and the network architecture
~\cite{kauffman,derrida86}. 
Moreover, distinct attractors in the configuration space are considered as corresponding 
to different cell types in a given organism and thus its scaling with 
the number of genes (nodes) across different organisms has been of great 
interest~\cite{kauffman,aldana05,bastolla98,bilke01,samuelsson03,drossel05,klemm05}.
Empirically, the number of cell types scales as the square-root
of the number of genes, and the same scaling relation was believed to 
hold between the number of attractors and the number of nodes in the 
Kauffman model at the critical point $p_c(K)$, supporting the hypothesis 
that living organisms should be between order and chaos~\cite{kauffman}.
Recently, however, it was found that under-sampling effects may hamper
numerical enumeration of distinct attractors~\cite{bastolla98,bilke01} and 
further investigations demonstrated that the total number of 
attractors grows faster than any power law with the 
system size~\cite{samuelsson03,drossel05}. On the other hand, 
it was also reported that attractors stable 
against deviation from synchronous update  
show a sub-linear scaling behavior~\cite{klemm05}.

Recent investiations of real gene networks suggest generalization of the 
original Kauffman model. First, the distribution of the regulating rules 
is structured 
showing a bias towards the canalyzing functions~\cite{harris02,kauffman03}. 
Second, the number of links or degree is not constant but different 
from node to node, resulting in broad degree distributions~\cite{albert02}. 
In the gene regulatory networks of {\it E. coli}~\cite{dobrin04,shenorr02,lee07} and 
yeast~\cite{guelzim02,tilee02,luscombe04,kauffman04,balcan05}, 
the distributions of out-degree 
(number of target genes for each regulator) and
in-degree (number of regulators for each target gene) were not  delta-functions 
but shown to take power-law or exponential-decaying form, respectively,  
although true asymptotic behaviors were hard to discern 
due to finite size effects.
While the effects of the 
structured distribution of regulating rules have been intensively studied
~\cite{kauffman03,kauffman04,moreira05}, it remains to show how the 
heterogeneous connectivity affects the dynamical stability
~\cite{oosawa02,aldana03}. 

We consider the Kauffman model on directed networks with general in- and 
out-degree distributions and compare two evolving dynamical configurations by 
computing their Hamming distance, to determine whether a given network is 
dynamically stable (zero distance) or unstable (non-zero distance) against 
perturbations. 
The critical point of the Boolean networks with power-law degree distributions 
was studied recently~\cite{aldana03}. In the present work, 
we show quantitatively 
how the Hamming distance behaves near the critical point, which 
will provide a deeper understanding of the critical phenomena of 
Boolean networks with heterogeneous connectivity patterns and 
insights into the interplay of structure and dynamics in living organisms.
The Hamming distance for infinite system size (thermodynamic limit) 
can be computed by the method presented in Ref.~\cite{lee07} and we 
here present a detailed description of the method along with 
a discussion on the effects of correlation between in- and out-degree 
of the same node.
Then, more importantly, we extend the method to derive the Hamming distance 
for finite system size, which enables us to check the analytic predictions 
with numerical simulation results.
Our main result is that for in-degree distributions with
a diverging second moment 
the Hamming distance increases very slowly 
when moving from the phase boundary towards the unstable phase and the width of 
the boundary in finite-size systems is very broad. This indicates that 
strongly heterogeneous genetic networks have a large capacity to stay at the 
edge of chaos when their structural and functional organization is subject to 
variation. 

The paper is organized as follows. 
We introduce the Kauffman model for Boolean networks in Sec.~\ref{sec:model}. 
In Sec.~\ref{sec:annealed}, the annealed approximation is described and used 
to compute the Hamming distance, which reveals different phases of the 
Boolean networks. The  finite-size effects on the critical phenomena of 
the Boolean networks are derived using the annealed approximation 
in Sec.~\ref{sec:fss}. Finally, the results are summarized 
and discussed in Sec.~\ref{sec:conclusion}. 

\section{Model and Hamming distance}
\label{sec:model}

In the Kauffman model, the dynamical configuration of $N$ Boolean 
variables at time $t$, 
$\Sigma(t) = \{\sigma_i(t)|i=1,2,\ldots,N\}$, is updated in parallel as
\begin{equation}
\sigma_i(t+1)=f_i(\Sigma_i(t)),
\end{equation}
where $\sigma_i$ for each $i$ takes $1$ or $0$ and 
$\Sigma_i(t) = \{\sigma_{i_1}(t), \sigma_{i_2}(t),\ldots, \sigma_{i_{k_i}}(t)\}$ denotes 
the configuration at time $t$ of the $k_i$ regulators  
$\reg_i=\{i_1, i_2, \ldots,i_{k_i}\}$, of 
the node $i$. The functional dependencies between nodes 
via $\{f_i(\Sigma_i)|i=1,2,\ldots,N\}$ constitute 
a directed network in which two nodes $i$ and $j$ 
are connected with a directed edge $(i, j)$ if $j\in\reg_i$, where $(i,j)$ is 
an outgoing edge of node $j$ and an incoming edge of node $i$. The quenched, i.e.,
 time-independent, regulating rules are random Boolean functions, i.e., they are 
chosen randomly such that $f_i(\Sigma_i)$ for a given $\Sigma_i$ is $1$ with 
probability $0\leq p\leq 1$ and $0$ with probability $1-p$. The parameter 
  $p$ deviating from $1/2$ indicates an asymmetry 
between expressed ($1$) and non-expressed ($0$) state of a gene. 
  
We focus on the following question: If one starts at time 
$t=0$ with two randomly chosen configurations, $\Sigma$ and $\hat{\Sigma}$ with 
$\hat{\sigma_j}\ne\sigma_j$ for all $j$, that is, all node states perturbed  
(altered), how many nodes remain perturbed at time $t>0$?  The fraction of 
these perturbed nodes or the Hamming distance between $\Sigma$ and $\hat{\Sigma}$ 
at time $t$ is defined as 
\begin{equation}
H(t)=\frac{1}{N}\sum_{i=1}^N \delta_{\sigma_i(t),1-\hat{\sigma_i}(t)}
\end{equation}
with $\delta_{a,b}$ being $1$ for $a=b$ and $0$ otherwise. The Hamming 
distance may vary between $0$ and $1$ depending on the dynamic asymmetry  
parameter $p$. We will see in the next section that 
the value of the Hamming distance in the stationary state 
may display a transition from zero to a 
non-zero value  as the network architecture and 
the parameter $p$ are varied. 

\section{Annealed approximation and phase transition of Boolean networks}
\label{sec:annealed}

In this section, we investigate the phase diagram of the Kauffman Boolean 
network defined in the previous section by computing analytically 
and numerically the Hamming distance for infinite system size. 
This allows us to understand different
phases of the Boolean networks determined by network structure and 
the parameter of dynamic asymmetry. 
Some of the results presented 
in this section are also found in Ref.~\cite{lee07}.

\subsection{Annealed approximation}

A recursion relation for the Hamming distance between consecutive time steps 
is obtained by the ``annealed'' approximation~\cite{derrida86}. While 
the regulation rule $f_i$ and the regulators $\reg_i$ are fixed for each 
node $i$ in the original model, one assigns them randomly to every node 
at every time step, keeping the in-degrees and the out-degrees, in the annealed 
approximation. Then the evolution of the Hamming distance $H_{k,q}(t)$ 
for the nodes with in-degree $k$ and out-degree $q$ is given by 
$
H_{k,q}(t+1) = \lambda [1-(1-\sum_{k',q'}  
q'P_d(k',q')H_{k',q'}(t)/\langle q\rangle )^k],
$
where $\lambda\equiv 2p(1-p)$, $P_d(k',q')$ is the joint distribution of $k'$ 
and $q'$, and $\langle q\rangle = \sum_{k,q} qP_d(k,q)$. 
The correlation between the degrees of neighboring nodes, 
$\{k,q\}$ and $\{k',q'\}$ is ignored in this formalism but 
will be discussed in Sec.~\ref{sec:conclusion}.
The parameter $\lambda$ ranging from $0$ to $1/2$ is the probability that 
$f_i$ yields different outputs for $\Sigma_i$ and $\hat{\Sigma}_i$ different 
and the term within the brackets represents the probability of the latter. Note 
that the degree distribution for the regulators is weighted by 
their out-degrees. If we introduce 
$\bar{H}(t) \equiv \sum_{k,q}[qP_d(k,q)/\langle q\rangle] H_{k,q}(t)$, it is 
obtained self-consistently and in turn $H(t)=\sum_{k,q} P_d(k,q) H_{k,q}(t)$ 
is computed as follows~\cite{lee07}:
\begin{eqnarray}
\bar{H}(t+1)&=& \lambda \sum_{k,q} \frac{qP_d(k,q)}{\langle q\rangle} [1-(1-\bar{H}(t))^k], \nonumber\\
H(t+1)&=& \lambda \sum_k P_d(k) [1-(1-\bar{H}(t))^k],
\label{eq:AA}
\end{eqnarray}
where $P_d(k)=\sum_q P_d(k,q)$ is the in-degree distribution. In the original 
Kauffman model where the in-degree is fixed to $k=K$, Eq.~(\ref{eq:AA}) reduces 
to $H(t+1)=\lambda [1-(1-H(t))^K]$~\cite{derrida86}.

\subsection{Ordered and chaotic phases}
\label{subsec:criticalpoint}

The limiting value $H = \lim_{t\to\infty} H(t)$ can characterize the 
system's response to dynamical perturbations. 
Replacing $H(t)$ and $\bar{H}(t)$ with $H$ and $\bar{H}$, respectively, and 
expanding the first line in Eq.~(\ref{eq:AA}) for small $\bar{H}$, 
one finds that $H=0$ and $\bar{H}=0$ 
for $\lambda <\lambda_c$ and $H>0$ and $\bar{H}>0$ 
for $\lambda>\lambda_c$, where the critical point $\lambda_c$ depends on 
the network topology via 
\begin{equation}
\lambda_c \equiv K^{-1} \quad \text{with} \quad K\equiv \sum_{k,q} \frac{kq P_{d}(k,q)}{\langle q\rangle}. 
\label{eq:lambdac}
\end{equation}
That is, the system is in the ordered phase, any perturbation making 
no effect on the system eventually, when $\lambda <\lambda_c$. 
On the other hand, 
the system does not remain in the same stationary state but 
shifts to another stationary state triggered by a perturbation when 
$\lambda > \lambda_c$.

\begin{figure}
\includegraphics[width=0.9\columnwidth]{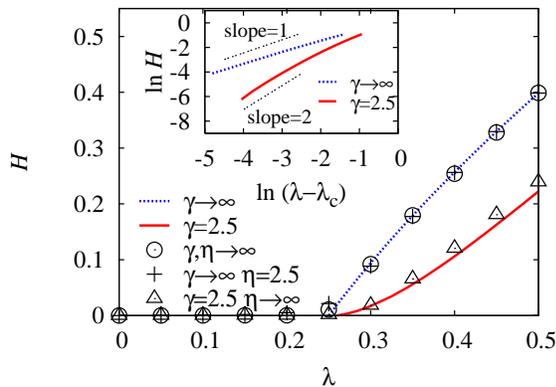}
\caption{(Color online) 
Hamming distance $H$ for the Kauffman model as a function of $\lambda$. The 
points are the simulation results for model networks~\cite{lee04npb} with 
$N=10^4$ and $\langle k\rangle=4$ that have $P_d(k,q)=g_\gamma(k)g_\eta(q)$, 
where $g_a(x)\sim x^{-a}$ for $a$ finite and  $g_\infty(x)= \langle k\rangle^x 
e^{-\langle k\rangle}/x!$. The Hamming distance $H$ 
was first averaged over time between $t=100$ and $t=200$, and then averaged 
over $1000$ different realizations of networks and regulating rules 
for which $H\ne 0$. The lines 
are the numerical solutions to Eq.~(\ref{eq:AA}) with $H(t)=\bar{H}(t)=H$ for 
all $t$ and the same degree distributions as in the simulation used.  (Inset) 
Plots of $\ln H$ versus  $\ln(\lambda-\lambda_c)$ with $\lambda_c = 0.25 $. 
The slopes agree with Eq.~(\ref{eq:beta}).} 
\label{fig:sn}
\end{figure}

The quantity $K$ may be considered as the average in-degree of regulators 
weighted by their out-degrees. 
Introducing the conditional average of out-degree, 
$q_k\equiv \sum_q q P_d(k,q)/P_d(k)$, we can rewrite the quantity 
$K$ as $K=\sum_k k (q_k/\langle q\rangle) P_d(k)$ and also the 
first relation in Eq.~(\ref{eq:AA})  as 
\begin{equation}
\bar{H}(t+1) = 
\lambda \sum_k \frac{q_k}{\langle q\rangle} P_d(k) [1-(1-\bar{H}(t))^k].
\label{eq:corr}
\end{equation}
If $q_k$ is independent of $k$ or (more strongly) 
  the in-degree and the out-degree 
of a node is not correlated statistically, 
it follows that $q_k=\langle q\rangle$ for all $k$, 
$K$ reduces to the conventional average in-degree $\langle k\rangle \equiv 
\sum_k k P_d(k)$, and $H(t)=\bar{H}(t)$. 
The analyses of the transcriptional regulatory networks of {\it E. coli} 
and yeast show no significant variation of $q_k$ with $k$~\cite{lee08} and 
so we will assume in the following that  $q_k=\langle q\rangle$ for all $k$. 
In Sec.~\ref{subsec:scaling}, we will discuss how our results for the 
critical phenomena would be changed by the $k$-dependence of $q_k$.
Under this assumption ($q_k=\langle q\rangle$), 
      the Hamming distance $H(t)$ depends 
only on the in-degree distribution $P_d(k)$ and the dynamics parameter 
$\lambda$.  

It has been shown that the scaling behavior of  
the average number of attractors with the system size 
remains to be the same 
for different out-degree distributions such as uniform, exponential, 
and power-law one~\cite{oosawa02}.
Our analysis based on the annealed approximation 
suggests further the irrelevance of the out-degree distribution 
to the Hamming distance.  
To confirm this as well as check the validity of the annealed 
approximation or Eq.~(\ref{eq:AA}), we 
performed simulations of the Kauffman model defined 
in Sec.~\ref{sec:model} on an ensemble of model networks constructed as follows~\cite{lee04npb}: 
i) Each of $N$ nodes has two indices $i_{\rm in}$ and $i_{\rm out}$, 
 which run from $1$ to $N$ respectively. The two indices are given independently to each node. 
ii) Choose a node $A$ with index $i_{\rm out}$ with probability $i_{\rm out}^{-\alpha_{\rm out}}/\sum_j j^{\alpha_{\rm out}}$. 
iii) Choose a node $B$ indexed $i_{\rm in}$ with probability $i_{\rm in}^{-\alpha_{\rm in}}/\sum_j j^{\alpha_{\rm in}}$. 
iv) Assign a link from the node $A$ to $B$ unless they are connected. 
v) Repeat ii) and iii) until the total number of links is $L$. 
The generated networks have $N$ nodes, $L$ links, and degree distribution given by 
$P_d(k,q)=P_d(k)P_d(q)$, where in-degree distribution takes the form 
$P_d(k)\sim k^{-\gamma}$ and the out-degree distribution takes the form 
$P_d(q)\sim q^{-\eta}$ with $\gamma= 1+1/\alpha_{\rm in}$ and $\eta = 1+1/\alpha_{\rm out}$. 
It is then obvious that $q_k=\langle q\rangle$ for all $k$.
When $\alpha_{\rm in}=0$ and thus all the nodes can have an incoming link with equal probability, 
the in-degree distribution becomes a Poisson one, $P_d(k) = \langle k\rangle^k e^{-\langle k\rangle}/k!$.
Thus the degree distribution may take  power-law or Poissonian form depending on 
the values of $\alpha_{\rm in}$ and $\alpha_{\rm out}$ 
corresponding to scale-free (SF) networks or completely random networks, respectively. 
The simulation results (data points) shown in Fig.~\ref{fig:sn} are compared with 
the numerical solutions (lines) to Eq.~(\ref{eq:AA}), the annealed approximation, which 
show a good agreement and support the validity of the annealed approximation. 
Also it is shown that the Hamming distance is the same for different out-degree distributions. 

The implication of Eq.~(\ref{eq:AA}) for SF networks has been discussed in 
Ref.~\cite{aldana03}, where $P_d(k)=k^{-\gamma}/\zeta(\gamma)$ 
for $k=1,2,\ldots$
with $\zeta(x)$ the Rieman-zeta function. Based on the result that 
$\langle k\rangle=\zeta(\gamma-1)/\zeta(\gamma)<2$ for $\gamma>2.47875\ldots$,
it was claimed~\cite{aldana03} that the abundance of SF networks with 
$2<\gamma<2.5$ in nature and society can be  attributed to the presence of both 
phases, stable and unstable, only in such networks. However, the values of 
$\langle k\rangle$ and $\gamma$ do not show such strong correlation in real 
networks. For instance, the average degree $\langle k\rangle$ ranges from 
$2.57$ (Internet router network) to $28.78$ (movie actor network) although the 
degree exponent $\gamma$ lies between $2$ and $3$~\cite{albert02}, which is 
possible due to the power-law behavior observed only asymptotically. 
We will show in the next section that 
root for the dynamical advantage of SF network lies elsewhere. 

\subsection{Critical exponents}
\label{subsec:criticalexponents}

Here we address the behavior of the Hamming distance around $\lambda_c$ 
for infinite system size. 
In that regime of $\lambda$, the Hamming distance is very small and its increase with $\lambda-\lambda_c$ 
can be characterized by a scaling exponent. This critical behavior of 
the Hamming distance is of interest to us because it shows how the network 
topology is related to the system's dynamic response.  

When the moments $\langle k^n \rangle = \sum_k P_d(k) k^n$ for 
all $n>0$ are finite, Eq.~(\ref{eq:AA}) can be written as 
\begin{equation}
H \simeq \lambda \sum_{n=1}^\infty \frac{(-1)^{n+1}}{n!} \langle k^n\rangle 
H^n.
\label{eq:Hexpand}
\end{equation}
Keeping the leading terms, we find that $H\simeq (\lambda/\lambda_c) H - 
\lambda \langle k^2\rangle H^2/2$, which  gives 
\[
H \sim \Delta
\]
with 
$\Delta \equiv \lambda/\lambda_c-1$ for $0<\Delta\ll 1$. 
This result can be represented as $H\sim \Delta^\beta$ with $\beta=1$, 
where we introduced the critical exponent $\beta$.  

On the other hand, if $P_d(k)\simeq c k^{-\gamma}$ for $k\gg 1$ with 
$c$ a constant, 
$\langle k^n \rangle$ diverges as $c k_{\rm max}^{n+1-\gamma}/(n+1-\gamma)$ for 
$n\geq \lceil \gamma\rceil - 1$ with $k_{\rm max}$ the largest 
in-degree and $\lceil x\rceil$ denoting the smallest integer not smaller than $x$. 
Applying the relation $\sum_{k>k_{\rm max}}P_d(k)\sim 1/N$ from 
the extreme value statistics~\cite{gumbel58}, one can see that 
$k_{\rm max}$ scales as $N^{1/(\gamma-1)}$.
The diverging terms in Eq.~(\ref{eq:Hexpand}) have alternative signs 
and lead to non-analytic terms in $H$ as described below~\cite{robinson51pr}. 
For small $H$, Eq.~(\ref{eq:AA}) reads as 
$H \simeq \lambda\sum_k P_d(k) [1-e^{-kH}]$ and recalling
the power-law form of $P_d(k)$, $P_d(k)\simeq c k^{-\gamma}$, 
we can utilize the fact that the Mellin transform of a function 
$F(\gamma,H) \equiv \sum_{k=1}^\infty k^{-\gamma} e^{-kH}$ is 
given by $\mathcal{F}(\gamma,s) = \int_0^\infty F(\gamma,H) H^{s-1}dH = 
\Gamma(s) \zeta(s+\gamma)$ with $\Gamma(x)$ the Gamma 
function~\cite{robinson51pr}. The inverse transform 
of $\mathcal{F}(\gamma,s)$ then is represented in terms of 
the poles of the Rieman-zeta function and the Gamma function, 
which gives $F(\gamma,H)=\int_{c-i\infty}^{c+i\infty} \mathcal{F}
(\gamma,s) H^{-s} ds = \Gamma(1-\gamma) H^{\gamma-1} + 
\sum_{n=0}^\infty (-1)^n /n! \zeta(\gamma-n) H^n$~\cite{robinson51pr}.
Therefore we find that, for $P_d(k)\simeq ck^{-\gamma}$~\cite{logarithmic},  
\begin{equation}
H \simeq \lambda \sum_{n=1}^{\lceil \gamma\rceil-2} 
\frac{(-1)^{n+1}}{n!} \langle k^n\rangle H^n 
-\lambda c \Gamma(1-\gamma) H^{\gamma-1} + \mathcal{O}(H^{\lceil \gamma\rceil -1}).
\label{eq:expand_singular}
\end{equation}
If $\gamma>3$, the $H^2$ term is the next leading term in the right-hand-side of 
Eq.~(\ref{eq:expand_singular}) and then the critical behavior is given by 
$H\sim \Delta$ as in the case of all $\langle k^n\rangle$ finite. 
On the other hand, 
if $2<\gamma<3$, the $H^{\gamma-1}$ term  becomes the next leading term and we find that   
$H \simeq (\lambda/\lambda_c) H - \lambda c \Gamma(1-\gamma) H^{\gamma-1}$.  
Therefore for $\lambda>\lambda_c$, 
\[H \sim \Delta^{1/(\gamma-2)}.
\] 
In summary, we can list the values of the critical exponent $\beta$ 
varying with the in-degree exponent $\gamma$ as~\cite{lee07} 
\begin{equation}
\beta = \left\{ 
\begin{array}{ll}
1 & (\gamma>3),\\
1/(\gamma-2) & (2<\gamma<3),
\end{array}
\right.
\label{eq:beta}
\end{equation}
which is confirmed numerically [See the inset of Fig.~\ref{fig:sn}]. 

\begin{figure}
\includegraphics[width=0.8\columnwidth]{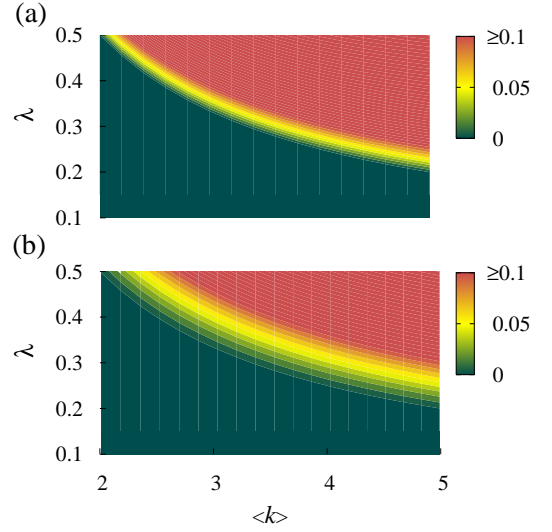}
\caption{(Color online) 
Phase diagram of the Kauffman model for (a) a Poisson in-degree distribution and 
(b) a power-law one with the exponent $\gamma=2.5$, both in case of uncorrelated 
in- and out-degree. The color encodes the Hamming distance $H$ obtained by 
numerically solving Eq.~(\ref{eq:AA}) under setting $H(t)=\bar{H}(t) = H$.}
\label{fig:pd}
\end{figure}

\begin{figure}
\includegraphics[width=0.66\columnwidth]{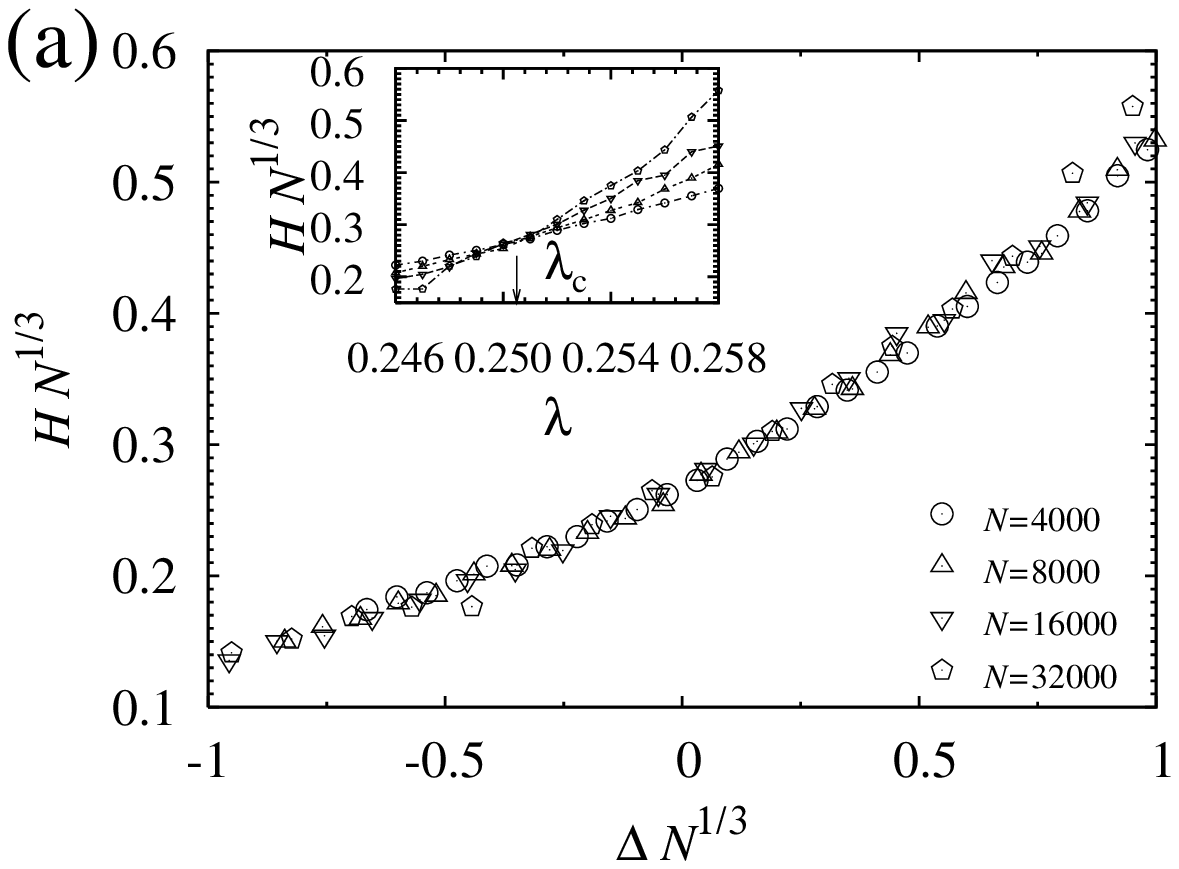}
\includegraphics[width=0.66\columnwidth]{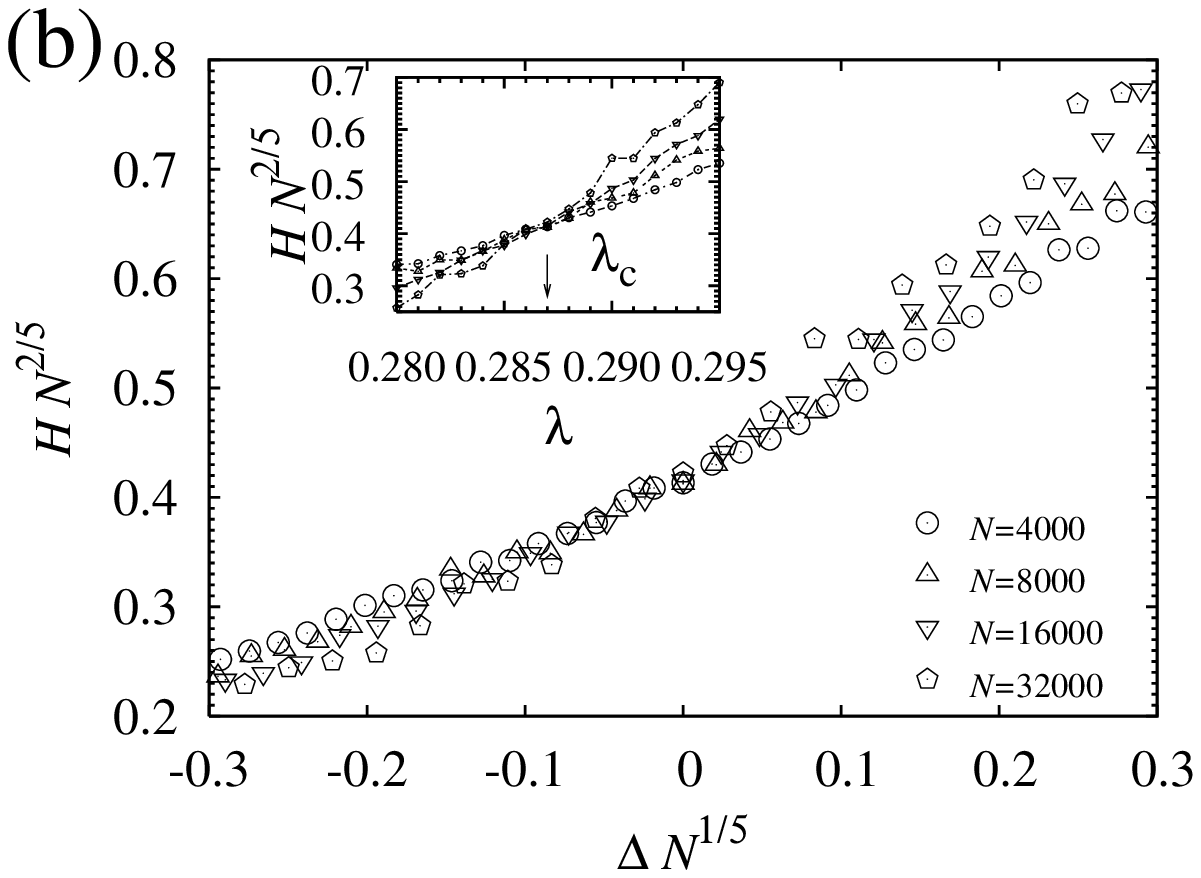}
\caption{
Finite-size scaling behavior in the critical regime. 
(a) Data collapse for different system sizes with 
$\gamma\to\infty$. The inset shows that $\lambda_c \simeq 0.2505(5)$. (b) Data 
collapse with $\gamma=2.5$ and $\lambda_c\simeq 0.287(1)$.} 
\label{fig:fss}
\end{figure}

The critical exponent $\beta$ varying with 
the in-degree distribution as in Eq.~(\ref{eq:beta}) 
is illuminating how the network topology affects the system's
response to perturbation. Figure~\ref{fig:pd} shows phase diagrams of 
the Kauffman model on model networks with a Poisson in-degree distribution and 
with a power-law in-degree distribution with the exponent $\gamma=2.5$, in 
which color represents the value of the Hamming distance. As shown in the 
figure, the SF networks with $2<\gamma<3$ and thus larger values of $\beta$ keep the 
Hamming distance non-zero but small in a much larger region in the 
$(\lambda,\langle k\rangle)$ plane than those with $\gamma>3$.
Structural and functional organization of cellular networks, 
parameterized here by $\langle k\rangle$ and $\lambda$ ($p$), respectively,
may be subject to unexpected changes. Our finding suggests that 
the systems with strong heterogeneous connectivity patterns can maintain 
their dynamic criticality robustly, and further, votes for 
the hypothesis that living organism's machinery lies at the edge of chaos. 

\section{Boolean networks of finite size}
\label{sec:fss}

In the thermodynamic limit $N\to\infty$, the critical point  
does not exhibit any dependence on the network topology.
However, for finite $N$, the critical point itself develops its dependence on the network 
topology: It is no more a point but has a non-zero width depending on $N$. 
Adopting the finite-size scaling ansatz~\cite{marroBook99}  
\begin{equation}
H = N^{-\beta/\mu} \ \Psi(\Delta N^{1/\mu})
\label{eq:fss}
\end{equation}
with the scaling function $\Psi(x)\to$ const. for $x\ll 1$ and 
$\Psi(x)\to x^\beta$ for $x\gg 1$, one can see that $H\sim N^{-\beta/\mu}$ in the 
{\it critical regime} $|\Delta N^{1/\mu}|\ll 1$.  In the $\lambda$ axis, this critical 
regime is $N^{-1/\mu}$ wide for $N$ finite and shrinks to zero in the thermodynamic limit. Therefore 
the scaling exponent $\mu$ describes the width of the critical regime for finite-size systems. 
In the critical regime, the cluster of perturbed nodes, explained below, exhibits scale invariance 
characterized by a power-law distribution of its size, which is connected to the 
behavior $H\sim N^{-\beta/\mu}$. We show in the next that the asymptotic behavior of the 
cluster size distribution can be derived using Eqs.~(\ref{eq:Hexpand}) and 
(\ref{eq:expand_singular}), which 
allows us to obtain the scaling exponent $\mu$ and to check Eq.~(\ref{eq:fss}).

\subsection{Evolution of perturbed-node clusters}
\label{subsec:clusters}

The parameter $\lambda$ denotes the probability that 
a node becomes perturbed ($\sigma_i(t+1) \ne \hat{\sigma_i}$)  
once the configuration of its neighbors are perturbed ($\Sigma_i \ne \hat{\Sigma_i}$). 
When $\lambda$ is zero, the Hamming distance becomes zero immediately even though all nodes 
were perturbed initially. As $\lambda$ increases from $0$, clusters appear, consisting of 
perturbed nodes that are connected by active edges. An edge from node $A$ to $B$ is inactive if 
the perturbation at node $A$ cannot bring any difference to the dynamical state of node $B$, and 
active otherwise. While a node's state can be totally irrelevant to its neighbor connected by 
an inactive edge, perturbation at one node can propagate to its neighbors through  active edges.   
The perturbed-node clusters evolve with time, decaying or growing.  
A perturbed node may become normal (non-perturbed) due to its regulators becoming normal at a certain 
time step, which leads to the decay of the cluster it belonged to. On the other hand,  normal nodes
may become perturbed due to its regulators becoming perturbed, which leads to the growth of a cluster.

When the parameter $\lambda$ reaches the critical regime or goes beyond it ($\lambda\gtrsim \lambda_c$), 
a giant cluster of perturbed nodes appears and contributes to the Hamming distance in the stationary state. 
There may exist many smaller clusters, but they are hard to survive eventually. 
A small cluster has a relatively small number of perturbed nodes, and they are surrounded 
by many normal nodes. Therefore the perturbed nodes in smaller clusters have higher 
chance of becoming normal than those in larger clusters, which leads to the higher chance of shrinking 
and decaying of smaller clusters. The perturbed nodes in the giant cluster, on the other hand,  
have more perturbed nodes as regulators and thus the giant cluster has a higher chance to survive; 
the probability becomes nonzero when $\lambda\gtrsim \lambda_c$. 
Therefore the Hamming distance $H$ can be approximated using the size of the largest cluster $S$ 
via $H\sim S/N$~\cite{aharony}. We will use this relation in the below to derive 
the asymptotic behavior of the size distribution of the clusters 
from the self-consistent equations for $H$, Eqs.~(\ref{eq:Hexpand}) and 
(\ref{eq:expand_singular}). 

\subsection{Cluster-size distribution in the annealed approximation}
\label{subsec:ps}

Let us denote the probability that a node belongs to a size-$s$ cluster (of 
perturbed nodes) by $P(s)$ and consider its generating function 
$\omega=\mathcal{P}(z)=\sum_s P(s)z^s$.
The Hamming distance is related to the generating function as 
\begin{equation}
H\simeq {S\over N}\simeq 
1-\sum_{s<S} P(s) \simeq \lim_{N\to\infty} [1-\mathcal{P}(z_N^*)],
\label{eq:HPz}
\end{equation} 
where 
$z_N^*=e^{-1/\tilde{S}}$ and $\tilde{S}$ satisfies $S_2\ll \tilde{S}\ll S$ with 
$S_2$ the second largest cluster size. This relationship, combined with 
Eqs.~(\ref{eq:Hexpand}) and (\ref{eq:expand_singular}), 
  gives us the functional form of 
the inverse function $z=\mathcal{P}^{-1}(\omega)$. Let's expand 
the inverse function around $\omega=1$ as  
$z=1-\sum_{n\geq 1} b_n (1-\omega)^n$.
Then one can see that this equation should reduce to 
Eqs.~(\ref{eq:Hexpand}) or  (\ref{eq:expand_singular}), depending on 
the in-degree distribution, with $z$ and $1-\omega$ 
replaced by $z_N^*$ and $H$, respectively, in the thermodynamic limit. 
Note that $z_N^*\to 1$ in the thermodynamic limit.
Therefore the inverse function $z=\mathcal{P}^{-1}(\omega)$ is expanded 
around $\omega=1$ as follows:
\begin{equation}
1-z \simeq (1-\lambda/\lambda_c) (1-\omega)  +
\lambda \langle k^2\rangle (1-\omega)^2 /2 + \cdots  
\label{eq:inverse1}
\end{equation}
for $\gamma>3$ and  
\begin{equation}
1-z \simeq (1-\lambda/\lambda_c) (1-\omega)  +
\lambda c \Gamma(1-\gamma) (1-\omega)^{\gamma-1} + \cdots 
\label{eq:inverse2}
\end{equation}
for $2<\gamma<3$,  where $c$ is the coefficient appearing in 
the asymptotic behavior of the in-degree distribution 
$P_d(k) \simeq ck^{-\gamma}$. 

The functional behavior of $\mathcal{P}(z)$ around $z=1$ is then derived 
from Eqs.~(\ref{eq:inverse1}) and (\ref{eq:inverse2}). 
At the critical point ($\lambda=\lambda_c$), it exhibits a 
singularity depending on the in-degree exponent:
\begin{equation}
1-\mathcal{P}(z)\sim \left\{
\begin{array}{ll}
(1-z)^{1/2} & (\gamma>3) \\
(1-z)^{1/(\gamma-1)} & (2<\gamma<3) 
\end{array}
\right.
\label{eq:sing}
\end{equation}
The asymptotic behavior of the cluster-size distribution $P(s)$ can be 
obtained from the functional form of $\mathcal{P}(z)$ since 
$P(s)=(1/s!) (d^s/dz^s)\mathcal{P}(z)|_{z=0}$. In particular, when 
$1-\mathcal{P}(z)\sim (1-z)^{\tau-1}$ with $\tau$ a non-integer, the 
cluster-size distribution takes a power-law form $P(s)\sim s^{-\tau}$. 
The asymptotic behavior of $P(s)$ is thus distinguished  
between $\gamma>3$ and $2<\gamma<3$ as follows:
\begin{equation}
P(s)\sim \left\{
\begin{array}{ll}
s^{-3/2} & (\gamma>3),\\ 
s^{-\gamma/(\gamma-1)} & (2<\gamma<3).
\end{array}
\right.
\label{eq:ps}
\end{equation} 
Scale invariance is typical of the system at the critical point and 
the power-law form of the cluster-size distribution  
is an example~\cite{aharony}. Our 
finding shows that the power-law exponent may vary with the degree exponent. 
In the subcritical ($\lambda<\lambda_c$) or supercritical ($\lambda>\lambda_c$) 
regime, the linear term in $1-\omega$ in 
Eqs.~(\ref{eq:inverse1}) and (\ref{eq:inverse2}) is dominant around $\omega=1$ and the cluster-size distribution 
takes an exponential-decaying form like $P(s) \sim e^{-s/s_c}$~\cite{lee04npb}.  
\subsection{Scaling exponents}
\label{subsec:scaling}

Once the asymptotic behavior of $P(s)$ is known, the scaling property of the 
largest cluster size may be derived.  
Using Eq.~(\ref{eq:ps}) in the relation $\sum_{s>S} P(s) \sim S/N$, 
a part of Eq.~(\ref{eq:HPz}), one can see 
that the size of the largest cluster $S$ 
scales with the system size $N$ as $S\sim N^{2/3}$ for $\gamma>3$ and 
$S\sim N^{(\gamma-1)/\gamma}$ for $2<\gamma<3$. 
It is known that the number of nonfrozen nodes  
in critical Boolean networks with any fixed number of 
inputs~\cite{kaufman05,mihaljev06} or 
fast-decaying in-degree distribution also scales as 
$N^{2/3}$~\cite{samuelsson06}. A node $i$ is called nonfrozen if 
its state ($\sigma_i$) is not fixed at either $0$ or $1$ 
in the stationary state and thus  may be perturbed (but not necessarily). 
The set of perturbed nodes is thus  a subset of the set of nonfrozen nodes 
and it is noteworthy that their sizes display the same scaling behavior.

The Hamming distance, $H\sim S/N$, in the critical regime is then given by 
\begin{equation}
H\sim \left\{
\begin{array}{ll}
N^{-1/3} & (\gamma>3), \\
N^{-1/\gamma} & (2<\gamma<3).
\end{array}
\right.
\label{eq:Hcritical}
\end{equation}
Since $H\sim N^{-\beta/\mu}$ in the critical regime from the finite-size 
scaling ansatz in Eq.~(\ref{eq:fss}), we find that the scaling 
exponent $\mu$ is given by 
\begin{equation}
\mu= \left\{ 
\begin{array}{ll}
3 & (\gamma>3),\\
\gamma/(\gamma-2) & (2<\gamma<3).
\end{array}
\right.
\label{eq:mu}
\end{equation}

To check numerically the derived finite-size scaling behavior of 
the Hamming distance,
we performed simulations of the Kauffman model on the model networks described 
in Sec.~\ref{subsec:criticalpoint} varying the system size. First, we 
plot the data of $HN^{\beta/\mu}$ as a function of  $\lambda$ 
for the same average degree ($\langle k\rangle=4$) and 
different system sizes ($N=4000, 8000, 16000, 32000$). They  
cross at one point, which determines the critical point 
$\lambda_c$ according to Eq.~(\ref{eq:fss}) 
(See the insets of Fig.~\ref{fig:fss}). 
Using these values of $\lambda_c$, we plotted the same data of 
$HN^{\beta/\mu}$ versus $\Delta N^{1/\mu}$ in Fig.~\ref{fig:fss}. 
The collapse of the data from different system sizes,  
while a slight deviation is seen in case of SF networks, 
presumably due to strong finite-size effects, 
 supports the scaling behavior of the 
Hamming distance in Eq.~(\ref{eq:fss}) with Eqs.~(\ref{eq:beta}) and 
(\ref{eq:mu}) used. 

The finite-size scaling behavior in Eq.~(\ref{eq:fss}) has been identified 
also in a wide range of dynamical systems on complex networks. 
In particular, the formation of a giant cluster in SF networks 
evolving by adding links  has been analyzed through its exact mapping to 
the $q=1$ Potts model and found to be characterized by 
$(\beta=1,\ \mu=3)$ for the degree exponent $\gamma  > 4$ and 
$(\beta=1/(\gamma-3),\ \mu=(\gamma-1)/(\gamma-3))$ for 
$3<\gamma<4$~\cite{lee04npb}.  
These exponents are very similar to Eqs.~(\ref{eq:beta}) and (\ref{eq:mu}).
Also in the Ising model~\cite{aleksiejuk02,leone02,igloi02,herrero04} and the Kuramoto 
model for synchronization phenomena~\cite{hong02,lee05}, the exponents are given by 
$(\beta=1/2,\ \mu=2)$ for the degree exponent $\gamma  > 5$ and 
$(\beta=1/(\gamma-3),\  \mu=(\gamma-1)/(\gamma-3))$ for $3<\gamma<5$. 
Such similar dependence of the scaling exponents on the degree exponent 
suggests a common framework to understand the critical phenomena on 
complex networks~\cite{igloi02,dorogovtsev02, hong07}. 

We note that while the scaling plot gives 
$\lambda_c \simeq 0.251$ for the completely random networks 
($\gamma\to\infty$) as predicted by  Eq.~(\ref{eq:lambdac}), 
the SF networks with $\gamma=2.5$ have $\lambda_c \simeq 
0.289$, deviating from the predicted value $0.25$. This 
deviation seems to be rooted in the use of the annealed approximation for  
the Hamming distance. It has been reported that 
the analytic prediction of the critical point 
in the framework of the mean-field theory 
deviates slightly from the numerical analysis in the Ising 
model~\cite{herrero04} and in the Kuramoto model 
for synchronization phenomena~\cite{lee05}. An improvement can be made by 
considering the Cayley tree with a given degree distribution as 
the underlying network topology~\cite{dorogovtsev02}. 

Our results show that the width of the critical regime  
$W\sim N^{-1/\mu}$ in the $\lambda$ axis increases as the in-degree exponent 
$\gamma$ decreases below $3$ while its scaling behavior remains the same 
for all $\gamma>3$. 
Since the number of genes in most organisms is not infinite but of 
order $10^5$ at most, the width of the critical regime may be between 
$\mathcal{O}(10^{-2})$ and $\mathcal{O}(1)$, depending on the in-degree exponent.  
Such a broad critical regime for small values of $\gamma$ 
should help living organisms to remain in the critical regime and in turn, 
to balance between robustness and evolvability.  
Individual dynamical responses depend  on the properties of the 
perturbed elements, 
i.e,, on their connectivities and regulating rules, which leads to perturbation 
propagation on various scales in heterogeneous networks~\cite{aldana03}.  Here 
we have analyzed the whole ensemble of such differentiated dynamical responses 
in heterogeneous networks and found that it can remain critical more easily 
with the help of extremely heterogeneous connectivity patterns.

The effects of correlation between in- and out-degree of 
the same node, which we have ignored so far, can be addressed in 
the results we obtained. In presence of the in- and out-degree 
correlation, $(q_k/\langle q\rangle)P_d(k)$ should be considered 
instead of $P_d(k)$ to compute the Hamming distance, as seen in 
Eq.~(\ref{eq:corr}). 
If it holds that  $q_k\sim k^{-\theta}$ for large $k$, 
we have to consider the effective in-degree exponent,  
$\gamma_{\rm eff} = \gamma+\theta$, in place of $\gamma$ for 
power-law in-degree distributions in all the results we obtained, 
including the scaling exponents in Eqs.~(\ref{eq:beta}) and (\ref{eq:mu}).

\section{Summary and Discussion}
\label{sec:conclusion}

In summary, we investigated the phase transition between the stable (ordered) 
and unstable (chaotic) phase in the Boolean dynamical network.  
Heterogeneous connectivities are found to broaden substantially 
the small Hamming distance region close to 
the phase boundary by suppressing the perturbation propagation in the 
unstable phase. Furthermore the transition region for finite system sizes 
turns out to be much wider than in homogeneous networks. 
Such a robust pseudo-criticality is expected 
to be also present in transcriptional regulatory networks and also 
in other biological networks such as neural networks~\cite{bornholdt03}, 
which can be a source for stability and  evolvability coexisting in living 
organisms.  

Our results suggest that the heterogeneous connectivity patterns of 
many biological networks have 
been selected in the course of evolution in part to serve for 
achieving stability 
and evolvability simultaneously. Therefore it would 
be desirable to propose a model in which 
heterogeneous connectivity patterns emerge driven 
by the evolutionary pressure towards a broad edge of chaos.
There are indeed models for co-evolution of network structure and 
dynamics, which reproduce networks 
with a selected number of links supporting dynamic 
criticality by dynamics-correlated addition and deletion 
of links~\cite{bornholdt00,liu06}. 
Similarly to these models, if one allows link 
rewiring only, preserving the total number of links, 
and make it happen depending on the network's dynamical state, 
the network is expected to be organized so as to have a broad degree 
distribution, which is under investigation.

While we focussed on the Hamming distance to capture the effects of 
structural features on the dynamic stability of Boolean networks, 
it would be also interesting to see how 
the heterogeneous connectivity patterns affect the properties of 
attractors in the configuration space, 
given the recent 
interests~\cite{bastolla98,bilke01,samuelsson03,drossel05,klemm05} 
in the scaling behavior of the number and length of the attractors in 
the critical Boolean networks. It has been shown that 
the median attractor length of SF networks  
is larger than that of networks with fixed in-degree at the critical 
point~\cite{iguchi07,kinoshita07}, but much more properties remain to 
be investigated.

The asymptotic behaviors of the in-degree distributions of real  
transcriptional regulatory networks of {\it E. coli}~\cite{shenorr02} 
and yeast~\cite{tilee02,luscombe04} are hard to 
discern due to finite size effects. In both organisms, there 
are about one hundred regulators (nodes with outgoing links), which 
impose a cut-off in the measured in-degree distributions. 
However, one can find for the network of yeast   
that its in-degree distribution is much broader than 
that of the networks generated by randomly rewiring the links, 
and that the Hamming distance of the Boolean dynamics is 
much slower than that in the randomized networks, 
demonstrating the contribution of heterogeneous connectivity pattern to 
maintaining dynamic criticality~\cite{lee07}.

It should be noted that the real transcriptional regulatory 
networks and other biological networks have much richer 
structural properties than described here  
and their relation to the dynamic criticality of the system 
is of interest. For example, the correlation of the degrees 
of neighboring nodes has been identified in many real-world 
networks~\cite{pastorsatorras01,newman02} including the yeast gene regulatory 
network~\cite{balcan05} and there are studies on the effects 
of the degree-degree correlation on the structure and dynamics of 
complex networks~\cite{vazquez03,bianconi06,noh07}. 
While it has been shown~\cite{noh07} that 
a negative (positive) degree-degree correlation  is irrelevant (relevant) 
to the percolation transition in complex networks, 
it still remains to be addressed 
how the correlation affects the critical phenomena of Boolean networks.


\begin{thebibliography}{19}

\expandafter\ifx\csname natexlab\endcsname\relax\def\natexlab#1{#1}\fi
\expandafter\ifx\csname bibnamefont\endcsname\relax
  \def\bibnamefont#1{#1}\fi
\expandafter\ifx\csname bibfnamefont\endcsname\relax
  \def\bibfnamefont#1{#1}\fi
\expandafter\ifx\csname citenamefont\endcsname\relax
  \def\citenamefont#1{#1}\fi
\expandafter\ifx\csname url\endcsname\relax
  \def\url#1{\texttt{#1}}\fi
\expandafter\ifx\csname urlprefix\endcsname\relax\def\urlprefix{URL }\fi
\providecommand{\bibinfo}[2]{#2}
\providecommand{\eprint}[2][]{\url{#2}}

\bibitem{thieffry98}
\bibinfo{author}{\bibfnamefont{D.}~\bibnamefont{Thieffry}},
  \bibinfo{author}{\bibfnamefont{A.~M.} \bibnamefont{Huerta}},
  \bibinfo{author}{\bibfnamefont{E.}~\bibnamefont{P\'{e}rez-Rueda}},
  \bibnamefont{and}
  \bibinfo{author}{\bibfnamefont{J.}~\bibnamefont{Collado-Vides}},
  \bibinfo{journal}{Bioessays} \textbf{\bibinfo{volume}{20}},
  \bibinfo{pages}{433} (\bibinfo{year}{1998}).

\bibitem{tilee02}
\bibinfo{author}{\bibfnamefont{T.~I.} \bibnamefont{Lee}}
  {\it et~al.}, \bibinfo{journal}{Science}
  \textbf{\bibinfo{volume}{298}}, \bibinfo{pages}{799} (\bibinfo{year}{2002}).

\bibitem{dobrin04}
\bibinfo{author}{\bibfnamefont{R.}~\bibnamefont{Dobrin}},
  \bibinfo{author}{\bibfnamefont{Q.~K.} \bibnamefont{Beg}},
  \bibinfo{author}{\bibfnamefont{A.-L.} \bibnamefont{Barab\'{a}si}},
  \bibnamefont{and} \bibinfo{author}{\bibfnamefont{Z.~N.}
  \bibnamefont{Oltvai}}, \bibinfo{journal}{BMC Bioinformatics}
  \textbf{\bibinfo{volume}{5}}, \bibinfo{pages}{10} (\bibinfo{year}{2004}).

\bibitem{guelzim02}
\bibinfo{author}{\bibfnamefont{N.}~\bibnamefont{Guelzim}},
  \bibinfo{author}{\bibfnamefont{S.}~\bibnamefont{Bottani}},
  \bibinfo{author}{\bibfnamefont{P.}~\bibnamefont{Bourgine}}, \bibnamefont{and}
  \bibinfo{author}{\bibfnamefont{F.}~\bibnamefont{K\'{e}p\`{e}s}},
  \bibinfo{journal}{Nature Genetics} \textbf{\bibinfo{volume}{31}},
  \bibinfo{pages}{60} (\bibinfo{year}{2002}).

\bibitem{shenorr02}
\bibinfo{author}{\bibfnamefont{S.~S.} \bibnamefont{Shen-Orr}},
  \bibinfo{author}{\bibfnamefont{R.}~\bibnamefont{Milo}},
  \bibinfo{author}{\bibfnamefont{S.}~\bibnamefont{Mangan}}, \bibnamefont{and}
  \bibinfo{author}{\bibfnamefont{U.}~\bibnamefont{Alon}},
  \bibinfo{journal}{Nature Genetics} \textbf{\bibinfo{volume}{31}},
  \bibinfo{pages}{64} (\bibinfo{year}{2002}).

\bibitem{babu04}
\bibinfo{author}{\bibfnamefont{M.~M.} \bibnamefont{Babu}}
{\it et al.,} \bibinfo{journal}{Curr. Opin. Struct. Biol} \textbf{\bibinfo{volume}{14}},
  \bibinfo{pages}{283} (\bibinfo{year}{2004}).

\bibitem{kauffman}
\bibinfo{author}{\bibfnamefont{S.}~\bibnamefont{Kauffman}},
  \bibinfo{journal}{J. Theor. Biol} \textbf{\bibinfo{volume}{22}},
  \bibinfo{pages}{437} (\bibinfo{year}{1969}); 
  \emph{\bibinfo{title}{The Origins of Order: Self-organization and Selection
  in Evolution}} (\bibinfo{publisher}{Oxford Univ. Press},
  \bibinfo{address}{Oxford}, \bibinfo{year}{1993}).
\bibitem{derrida86}
\bibinfo{author}{\bibfnamefont{B.}~\bibnamefont{Derrida}} \bibnamefont{and}
  \bibinfo{author}{\bibfnamefont{Y.}~\bibnamefont{Pomeau}},
  \bibinfo{journal}{Europhys. Lett.} \textbf{\bibinfo{volume}{1}},
  \bibinfo{pages}{45} (\bibinfo{year}{1986}).

\bibitem{aldana05}
M. Aldana, S. Coppersmith, and L.P. Kadanoff, in 
{\it Perspectives and Problems in Nonlinear Science. 
 A celebratory volule in honor of Lawrence Sirovich}. 
 Springer Applied Mathematical Sciences Series. E. Kaplan, J.E. Marsden, and K.R. Sreenivasan Eds. (2003). 

\bibitem{bastolla98}
U. Bastolla and G. Parisi, Physica D {\bf 115}, 203 (1998); 219 (1998).

\bibitem{bilke01}
\bibinfo{author}{\bibfnamefont{S.}~\bibnamefont{Bilke}} \bibnamefont{and}
  \bibinfo{author}{\bibfnamefont{F.}~\bibnamefont{Sjunnesson}},
  \bibinfo{journal}{Phys. Rev. E} \textbf{\bibinfo{volume}{65}},
  \bibinfo{pages}{016129} (\bibinfo{year}{2001}).

\bibitem{samuelsson03}
\bibinfo{author}{\bibfnamefont{B.}~\bibnamefont{Samuelsson}} \bibnamefont{and}
  \bibinfo{author}{\bibfnamefont{C.}~\bibnamefont{Troein}},
  \bibinfo{journal}{Phys. Rev. Lett.} \textbf{\bibinfo{volume}{90}},
  \bibinfo{pages}{098701} (\bibinfo{year}{2003}).

\bibitem{drossel05}
\bibinfo{author}{\bibfnamefont{B.}~\bibnamefont{Drossel}},
\bibinfo{author}{\bibfnamefont{T.}~\bibnamefont{Mihaljev}},
  \bibnamefont{and}
  \bibinfo{author}{\bibfnamefont{F.}~\bibnamefont{Greil}},
  \bibinfo{journal}{Phys. Rev. Lett.} \textbf{\bibinfo{volume}{94}},
  \bibinfo{pages}{088701} (\bibinfo{year}{2005}).

\bibitem{klemm05}
\bibinfo{author}{\bibfnamefont{K.}~\bibnamefont{Klemm}} \bibnamefont{and}
  \bibinfo{author}{\bibfnamefont{S.}~\bibnamefont{Bornholdt}},
  \bibinfo{journal}{Phys. Rev. E} \textbf{\bibinfo{volume}{72}},
  \bibinfo{pages}{055101(R)} (\bibinfo{year}{2005}).




\bibitem{harris02}
\bibinfo{author}{\bibfnamefont{S.~E.} \bibnamefont{Harris}},
  \bibinfo{author}{\bibfnamefont{B.~K.} \bibnamefont{Sawhill}},
  \bibinfo{author}{\bibfnamefont{A.}~\bibnamefont{Wuensche}}, \bibnamefont{and}
  \bibinfo{author}{\bibfnamefont{S.}~\bibnamefont{Kauffman}},
  \bibinfo{journal}{Complexity} \textbf{\bibinfo{volume}{7}},
  \bibinfo{pages}{23} (\bibinfo{year}{2002}).

\bibitem{kauffman03}
\bibinfo{author}{\bibfnamefont{S.}~\bibnamefont{Kauffman}},
  \bibinfo{author}{\bibfnamefont{C.}~\bibnamefont{Peterson}},
  \bibinfo{author}{\bibfnamefont{B.}~\bibnamefont{Samuelsson}},
  \bibnamefont{and} \bibinfo{author}{\bibfnamefont{C.}~\bibnamefont{Troein}},
  \bibinfo{journal}{Proc. Natl. Acad. Sci. U.S.A.}
  \textbf{\bibinfo{volume}{100}}, \bibinfo{pages}{14796}
  (\bibinfo{year}{2003}).
\bibitem{albert02}
R. Albert and A.-L. Barab\'{a}si, Rev. Mod. Phys. {\bf 74}, 47 (2002)
\bibitem{lee07}
\bibinfo{author}{\bibfnamefont{D.-S.} \bibnamefont{Lee}} \bibnamefont{and}
  \bibinfo{author}{\bibfnamefont{H.}~\bibnamefont{Rieger}},
  \bibinfo{journal}{J. Theor. Biol.}
  \textbf{\bibinfo{volume}{248}}, \bibinfo{pages}{618}
  (\bibinfo{year}{2007}).

\bibitem{kauffman04}
\bibinfo{author}{\bibfnamefont{S.}~\bibnamefont{Kauffman}},
  \bibinfo{author}{\bibfnamefont{C.}~\bibnamefont{Peterson}},
  \bibinfo{author}{\bibfnamefont{B.}~\bibnamefont{Samuelsson}},
  \bibnamefont{and} \bibinfo{author}{\bibfnamefont{C.}~\bibnamefont{Troein}},
  \bibinfo{journal}{Proc. Natl. Acad. Sci. U.S.A.}
  \textbf{\bibinfo{volume}{101}}, \bibinfo{pages}{17102}
  (\bibinfo{year}{2004}).

\bibitem{balcan05}
\bibinfo{author}{\bibfnamefont{D.}~\bibnamefont{Balcan}},
  \bibinfo{author}{\bibfnamefont{A.}~\bibnamefont{Kabak\c{c}io\u{g}lu}},
  \bibinfo{author}{\bibfnamefont{M.}~\bibnamefont{Mungan}},
  \bibnamefont{and} \bibinfo{author}{\bibfnamefont{A.}~\bibnamefont{Erzan}},
  \bibinfo{journal}{PLoS One}
  \textbf{\bibinfo{volume}{2}}, \bibinfo{pages}{e501}
  (\bibinfo{year}{2005}).


\bibitem{luscombe04}
N.M. Luscombe, M.M. Babu, H. Yu, M. Snyder, S.A. Teichmann, and M. Gerstein, 
Nature {\bf 431}, 308 (2004).

\bibitem{moreira05}
\bibinfo{author}{\bibfnamefont{A.~A.}~\bibnamefont{Moreira}} \bibnamefont{and}
  \bibinfo{author}{\bibfnamefont{L.~A.~N.} \bibnamefont{Amaral}},
  \bibinfo{journal}{Phys. Rev. Lett.} \textbf{\bibinfo{volume}{94}},
  \bibinfo{pages}{218702} (\bibinfo{year}{2005}).

\bibitem{oosawa02}
\bibinfo{author}{\bibfnamefont{C.}~\bibnamefont{Oosawa}} \bibnamefont{and}
  \bibinfo{author}{\bibfnamefont{M.~A.} \bibnamefont{Savageau}},
  \bibinfo{journal}{Phyisca D} \textbf{\bibinfo{volume}{170}},
  \bibinfo{pages}{143} (\bibinfo{year}{2002}).

\bibitem{aldana03}
\bibinfo{author}{\bibfnamefont{M.}~\bibnamefont{Aldana}} \bibnamefont{and}
  \bibinfo{author}{\bibfnamefont{P.}~\bibnamefont{Cluzel}},
  \bibinfo{journal}{Proc. Natl. Acad. Sci. U.S.A.}
  \textbf{\bibinfo{volume}{100}}, \bibinfo{pages}{8710} (\bibinfo{year}{2003}).
\bibitem{lee08}
D.-S. Lee, unpublished data (2008).

\bibitem{lee04npb}
\bibinfo{author}{\bibfnamefont{D.-S.} \bibnamefont{Lee}},
  \bibinfo{author}{\bibfnamefont{K.-I.} \bibnamefont{Goh}},
  \bibinfo{author}{\bibfnamefont{B.}~\bibnamefont{Kahng}}, \bibnamefont{and}
  \bibinfo{author}{\bibfnamefont{D.}~\bibnamefont{Kim}},
  \bibinfo{journal}{Nucl. Phys. B} \textbf{\bibinfo{volume}{696}},
  \bibinfo{pages}{351} (\bibinfo{year}{2004}).

  \bibitem{gumbel58}
\bibinfo{author}{\bibfnamefont{E.J.}~\bibnamefont{Gumbel}} 
  \emph{\bibinfo{title}{Statistics of Extremes}}
  (\bibinfo{publisher}{Columbia Univ. Press}, \bibinfo{address}{New York},
  \bibinfo{year}{1958}).


\bibitem{robinson51pr}
\bibinfo{author}{\bibfnamefont{J.E.} \bibnamefont{Robinson}},
  \bibinfo{journal}{Phys. Rev.} \textbf{\bibinfo{volume}{83}},
  \bibinfo{pages}{678} (\bibinfo{year}{1951}).

\bibitem{logarithmic}
There is a logarithmic term $H^{\gamma-1}\ln H$ in Eq.~(\ref{eq:expand_singular}) in case of 
$\gamma$ an integer~\cite{robinson51pr}. 

\bibitem{marroBook99}
\bibinfo{author}{\bibfnamefont{J.}~\bibnamefont{Marro}} \bibnamefont{and}
  \bibinfo{author}{\bibfnamefont{R.}~\bibnamefont{Dickman}},
  \emph{\bibinfo{title}{Nonequilibrium Phase Transitions in Lattice Models}}
  (\bibinfo{publisher}{Cambridge Univ. Press}, \bibinfo{address}{Cambridge},
  \bibinfo{year}{1999}).

  \bibitem{aharony}
\bibinfo{author}{\bibfnamefont{D.}~\bibnamefont{Stauffer}} \bibnamefont{and}
  \bibinfo{author}{\bibfnamefont{A.}~\bibnamefont{Aharony}},
  \emph{\bibinfo{title}{Introduction to percolation theory}}
  (\bibinfo{publisher}{Taylor \& Francis}, \bibinfo{address}{London},
  \bibinfo{year}{1994}).


\bibitem{kaufman05}
\bibinfo{author}{\bibfnamefont{V.}~\bibnamefont{Kaufman}},
\bibinfo{author}{\bibfnamefont{T.}~\bibnamefont{Mihaljev}},
  \bibnamefont{and}
  \bibinfo{author}{\bibfnamefont{B.} \bibnamefont{Drossel}},
  \bibinfo{journal}{Phys. Rev. E} \textbf{\bibinfo{volume}{72}},
  \bibinfo{pages}{046124} (\bibinfo{year}{2005}).

\bibitem{mihaljev06}
\bibinfo{author}{\bibfnamefont{T.}~\bibnamefont{Mihaljev}},
  \bibnamefont{and}
  \bibinfo{author}{\bibfnamefont{B.} \bibnamefont{Drossel}},
  \bibinfo{journal}{Phys. Rev. E} \textbf{\bibinfo{volume}{74}},
  \bibinfo{pages}{046101} (\bibinfo{year}{2006}).

\bibitem{samuelsson06}
\bibinfo{author}{\bibfnamefont{B.}~\bibnamefont{Samuelsson}},
  \bibnamefont{and}
  \bibinfo{author}{\bibfnamefont{J.E.} \bibnamefont{Socolar}},
  \bibinfo{journal}{Phys. Rev. E} \textbf{\bibinfo{volume}{74}},
  \bibinfo{pages}{036113} (\bibinfo{year}{2006}).

 \bibitem{aleksiejuk02}
\bibinfo{author}{\bibfnamefont{A.}~\bibnamefont{Aleksiejuk}},
\bibinfo{author}{\bibfnamefont{J.A.}~\bibnamefont{Holyst}},
  \bibnamefont{and}
\bibinfo{author}{\bibfnamefont{D.}~\bibnamefont{Stauffer}},
  \bibinfo{journal}{Physica A} \textbf{\bibinfo{volume}{310}},
  \bibinfo{pages}{260} (\bibinfo{year}{2002}).

  \bibitem{leone02}
\bibinfo{author}{\bibfnamefont{M.}~\bibnamefont{Leone}},
\bibinfo{author}{\bibfnamefont{A.}~\bibnamefont{V\'{a}zquez}},
\bibinfo{author}{\bibfnamefont{A.}~\bibnamefont{Vespignanai}},
  \bibnamefont{and}
  \bibinfo{author}{\bibfnamefont{R.} \bibnamefont{Zecchina}},
  \bibinfo{journal}{Eur.Phys.J.B} \textbf{\bibinfo{volume}{28}},
  \bibinfo{pages}{191} (\bibinfo{year}{2002}).

\bibitem{igloi02}
\bibinfo{author}{\bibfnamefont{F.}~\bibnamefont{Igl\'{o}i}},
  \bibnamefont{and}
\bibinfo{author}{\bibfnamefont{L.}~\bibnamefont{Turban}},
  \bibinfo{journal}{Phys. Rev. E} \textbf{\bibinfo{volume}{66}},
  \bibinfo{pages}{036140} (\bibinfo{year}{2002}).

\bibitem{herrero04}
\bibinfo{author}{\bibfnamefont{C.P.}~\bibnamefont{Herrero}},
  \bibinfo{journal}{Phys. Rev. E} \textbf{\bibinfo{volume}{69}},
  \bibinfo{pages}{067109} (\bibinfo{year}{2004}).

  \bibitem{hong02}
\bibinfo{author}{\bibfnamefont{H.}~\bibnamefont{Hong}},
\bibinfo{author}{\bibfnamefont{M.Y.}~\bibnamefont{Choi}},
  \bibnamefont{and}
\bibinfo{author}{\bibfnamefont{B.J.}~\bibnamefont{Kim}},
  \bibinfo{journal}{Phys. Rev. E} \textbf{\bibinfo{volume}{65}},
  \bibinfo{pages}{026139} (\bibinfo{year}{2002}).

\bibitem{lee05}
\bibinfo{author}{\bibfnamefont{D.-S.} \bibnamefont{Lee}},
  \bibinfo{journal}{Phys. Rev. E} \textbf{\bibinfo{volume}{72}},
  \bibinfo{pages}{026208} (\bibinfo{year}{2005}).


\bibitem{dorogovtsev02}
\bibinfo{author}{\bibfnamefont{S.N.}~\bibnamefont{Dorogovtsev}},
  \bibinfo{author}{\bibfnamefont{A.V.} \bibnamefont{Goltsev}},
  \bibnamefont{and}
  \bibinfo{author}{\bibfnamefont{J.F.F.} \bibnamefont{Mendes}},
  \bibinfo{journal}{Phys. Rev. E} \textbf{\bibinfo{volume}{66}},
  \bibinfo{pages}{016104} (\bibinfo{year}{2002}).


\bibitem{hong07}
\bibinfo{author}{\bibfnamefont{H.}~\bibnamefont{Hong}},
  \bibinfo{author}{\bibfnamefont{M.} \bibnamefont{Ha}},
  \bibnamefont{and}
  \bibinfo{author}{\bibfnamefont{H.}\bibnamefont{Park}},
  \bibinfo{journal}{Phys. Rev. Lett.} \textbf{\bibinfo{volume}{98}},
  \bibinfo{pages}{258701} (\bibinfo{year}{2007}).


\bibitem{bornholdt03}
S. Bornholdt and T. R\"{o}hl, Phys. Rev. E {\bf 67}, 066118 (2003).

\bibitem{bornholdt00}
S. Bornholdt and T. Rohlf, Phys. Rev. Lett. {\bf 84}, 6114 (2000).

\bibitem{liu06}
M. Liu and K.E. Bassler, Phys. Rev. E {\bf 74}, 041910 (2006).

\bibitem{iguchi07}
K. Iguchi, S. Kinoshita, and H.S. Yamamda, J. Theor. Biol. {\bf 247}, 138 (2007).

\bibitem{kinoshita07}
S. Kinoshita, K. Iguchi, and H.S. Yamamda, 
AIP Conference Proceedings {\bf 982}, 2128 (2008).
\bibitem{pastorsatorras01}
R. Pastor-Satorras, A. Vazquez, and A. Vespignani, Phys. Rev. Lett. 
{\bf 87}, 258701 (2001).

\bibitem{newman02}
M.E.J. Newman, Phys. Rev. Lett. {\bf 89}, 208701 (2002).

\bibitem{vazquez03}
A. Vazquez and Y. Moreno, Phys. Rev. E {\bf 67}, 015101(R) (2003).

\bibitem{bianconi06}
G. Bianconi and M. Marsili, Phys. Rev. E {\bf 76}, 026116 (2007).

\bibitem{noh07}
J.D. Noh, Phys. Rev. E {\bf 76}, 026116 (2007).

\end{thebibliography}
\end{document}